\documentclass[aps,pre,twocolumn,10pt]{revtex4-2}

\usepackage{graphicx}  
\usepackage{dcolumn}   
\usepackage{bm}        
\usepackage[utf8]{inputenc}
\usepackage[T1]{fontenc}
\usepackage{amsmath}

\usepackage{hyperref}
\usepackage{nameref}
\usepackage{cleveref}
\hypersetup{colorlinks,linkcolor={black},citecolor={blue}, urlcolor={blue}}
\usepackage{booktabs}

\begin{document}

\title{Stability and Dynamics of Skyrmion and Skyrmion Bags Explored under the Influence of Out-of-Plane Strain and Its Gradient}

\author{Naim Ahmad}
\author{Chirag Kalouni}
\author{Abhay Singh Rajawat}
\author{Waseem Akhtar}
\altaffiliation{author who may correspond:\\ {makhtar5@jmi.ac.in}}  
\affiliation{Department of Physics, Jamia Millia Islamia, New Delhi-110025}

\begin{abstract}
Skyrmions as well as skyrmion bags in magnetic thin films are promising candidates for future high-density memory devices. The observation of skyrmion bags in liquid crystals and their predicted existence in ferromagnetic films has sparked theoretical studies on current-induced dynamics of these topological charges. Here, using micromagnetism, we study the impact of out-of-plane strain on the stability of skyrmion and skyrmion bags in ferromagnetic thin films. We further investigate the current-induced dynamics in the presence of an out-of-plane strain gradient. We demonstrate that the out-of-plane strain gradient direction with respect to the electron flow can be an efficient way to control the dynamics of topological charges. Specifically, the deflection of skyrmion bags correlates with their topological degree, and an appropriate strain gradient can counteract the skyrmion Hall effect, enabling straight-line movement. Our micromagnetic simulations align well with theoretical predictions from the Thiele equation.
\end{abstract}

\maketitle  

\section{Introduction }

Magnetic skyrmions are topologically protected chiral spin textures experimentally observed in variety of  material system \cite{1,2,3,4,5,6,7,8,9,10,11,12,13,68}.Room temperature nucleation and its efficient current induced dynamics governed by the interfacial interaction in technologically friendly materials have made magnetic skyrmions appealing for room temperature (RT) memory applications \cite{14,15,16,17,18}.The existence or absence of skyrmions serves as the binary bit representation in race track memory \cite{19,20,21,22,23,24,25,26,34},however it necessitates highly precise spacing of the skyrmion in order to express high density memory. Recently, skyrmion bags \cite{27,28,29,30,31,32,33,34} which consist of an outer boundary-defining skyrmion hosting several interior skyrmions with opposite topological charge, have attracted significant interest as it not only has all the functionalities of topological protection and efficient current induced-dynamics associated with skyrmion, but also enable higher density packing of magnetic information as compared to individual skyrmions in magnetic system. However, stabilization of skyrmion bags in magnetic thin film still remain elusive. The Skyrmion hall effect (SkHE) \cite{35,36,37,38,39} associated with the current induced dynamics of these spin textures, is one of the constraints to the practical use of skyrmions or skyrmion bags in racetrack memory devices. The Magnus force \cite{36,41}due to the topological charge of skyrmion/skyrmion-bags leads to the deflection of these spin-textures towards the boundary of the nano track. This effect increases the risk of annihilation of topological charges at the edges and hence data loss, making it difficult to achieve reliable and stable data storage.Stabilizing skyrmion in magnetic thin films and overcoming the Skyrmion Hall Effect (SkHE) are indeed significant challenges for the practical applications. Key Strategies to mitigate SkHE in magnetic thin films includes anisotropy gradients \cite{44,45,46,47,48}, Magnetization Gradient \cite{39},temperature gradients \cite{62,63}, constrained geometries \cite{64}, and variations in nanotrack width \cite{61}.Strain engineering is another highly promising and energy-efficient method for manipulating skyrmions in magnetic materials \cite{54,59,49,50,51,60,53,54,55,57,46.1,58}. Here, in this paper, we have performed micromagnetic simulations to study the effect of out-of-plane strain and its gradient on the stability and current induced dynamics, respectively, for skyrmion and skyrmion bags in magnetic thin films. The topological charge Q for a spin texture is defined as:
\begin{equation}
    Q = \frac{1}{4\pi} \int \textbf{m} \cdot \left( \frac{\partial \textbf{m}}{\partial x} \times \frac{\partial \textbf{m}}{\partial y} \right) \, dx \, dy          
\end{equation}
where $\textbf{m}$ is the unit vector of the magnetization.In this paper, we define skyrmion bag S(N) as an outer Q = +1 skyrmion hosting N number of Q = -1 skyrmion within.Thus, S (2) represents a spin texture with an outer Q = +1 skyrmion nesting two inner skyrmions with Q = -1, and so on.  Our simulations clearly demonstrates that the stability of a skyrmion or skyrmion bag is enhanced in the presence of an out-of- plane strain $(\epsilon_{zz})$. Furthermore, we study the effect of out-of-plane strain-gradient  $(\partial\epsilon_{zz})$ on the current induced dynamics of skyrmion and skyrmion bags. We found that the strain gradient gives a very efficient way of controlling the SkHE depending upon whether the gradient is applied parallel or perpendicular to the direction of electron flow (\textbf{j}$_\textbf{e}$). Our results demonstrate that strain gradient has the ability to tune dynamics of topological charges that could lead to the creation of highly efficient and reliable memory devices.The advantage of using strain is that it does not require any additional magnetic field or current, which can cause additional energy dissipation and heat generation.Strain can be locally applied, allowing for more precise control over the motion of magnetic spin textures.
\section{Modelling and Simulation}
In this paper, we have performed micromagnetic simulations using MuMax3 \cite{67} , a GPU-accelerated micromagnetic simulation package. The magnetization dynamics in the absence of spin polarized current is given by the Landau-Lifshitz-Gilbert (LLG) equation given as:
 \begin{equation} 
 \frac{\partial \textbf{m}}{\partial t} = -\gamma \, \textbf{m} \times \textbf{H}_{\text{eff}} + \alpha \, \textbf{m} \times \frac{\partial \textbf{m}}{\partial t}
  \end{equation}
where $\gamma$ is the gyromagnetic ratio, $\textbf{m}$ is the unit magnetization vector, $\alpha$ is the Gilbert damping coefficient and $ \textbf{H}_{eff} $ is the effective magnetic field given as: 
\begin{equation}
    \textbf{H}_{\text{eff}} = -\frac{1}{\mu_{0} M_{\text{sat}}} \frac{\partial W}{\partial \textbf{m}}\nonumber
\end{equation}

The magnetic energy density W consist of the perpendicular magnetic anisotropy (PMA), the Heisenberg exchange, the Dzyaloshinskii–Moriya interaction (DMI), the demagnetization field and the uniaxial strain energy.
First part of the simulation was to study the stability of the topological charges in a ferromagnetic layer by calculating the total micromagnetic energy in the presence of out-of-plane uniaxial strain $(\epsilon_{zz})$ , where strain energy term was defined as
  $W_{strain}= -\frac{3}{2} \epsilon k \lambda  (\textbf{m} \cdot \hat{\textbf{z}})^2 $  \ \cite{52}. Here  $k$ is the young’s modulus, $\lambda$ is the magnetostriction constant and $ \hat{\textbf{z}}$  is the direction of strain. In this part of the simulation, we defined 400 $\times$ 400 nm$^2$  square film of thickness 0.4 nm and used a cell size measuring 2 $\times$ 2 $\times$ 0.4 nm$^3$. Our objective was to stabilize the Neel skyrmion and skyrmion bags at the center of the film.
  For stabilization of skyrmion  with Q = $+$1 or Q = $-$1 in a film we define a neel skyrmion at the center of the film whose core polarization is along $+z$ and $-z$ direction respectively and allow this configuration to relax in the presence of user defined material parameters given below.
For a skyrmion bag S(N), we consider a ferromagnetic thin film with magnetization oriented along the $-z$ direction. At the center of this film, we define a circular disk with a radius of 100 nm and a height of 0.4 nm, featuring magnetization aligned in the $+z$ direction. Within this central region, we introduce N smaller circular disks, each with a radius of 10 nm and a height of 0.4 nm, whose magnetization is directed along the $-z$ direction. This initial configuration is allowed to relax for the formation of desired skyrmion bag S(N).

The second part of the simulation involves the study of current induced dynamics for different topological charges in the presence of out-of-plane strain gradient $(\partial \epsilon_{zz})$ along the x or y direction of the film/nanostrip. Here the in-plane spin polarized current is flowing through the ferromagnetic layer, thus the dynamics of skyrmion/skyrmion bags are governed by the extended LLG equation containing the Zhang-Li spin transfer torque (STT) \cite{65} term as following:$$\frac{\partial \textbf{m}}{\partial t}=\frac{1}{(1+\alpha^2 )}((1+\beta \alpha)  \textbf{m} \times (\textbf{m} \times \textbf{u} \cdot \nabla )\textbf{m}$$
\begin{equation} 
   +(\beta-\alpha) \textbf{m}  \times (\textbf{u} \cdot \nabla) \textbf{m} )  
\end{equation} where $ \textbf{u}=\frac{\mu_B \mu_o \textbf{j}}{2e\gamma_o M_{sat} (1+\beta^2)} $ is spin drift velocity defined for a current density $\textbf{j}$ , $\mu_B$ is Bohr magneton, $\mu_0$ is permeability,e is charge of electron, $\beta$ is the non adiabaticity parameter, $\alpha$ is gyrocoupling damping parameter and $M_{sat}$ is saturation magnetization in Tesla.  The linear strain gradient along the x and y axis i.e., parallel and perpendicular to the direction of electron flow is applied by varying strain gradient in the strain energy term $(W_{strain})$ as $\epsilon_{zz}(i)=\epsilon_0+\partial\epsilon_{zz} \cdot i$. Here choosing $i$ = $x$ or $y$ defines the direction of strain gradient parallel or perpendicular, respectively with respect to the applied current direction which is chosen to be along $-x$ direction through out the manuscript. For current induced dynamics of skyrmion we have taken a large grid size of 512$\times$256$\times$1 with the same cell size of 2 $\times$2$\times$0.4 $nm^3$. 
\\
\begin{figure*}[t]
  \centering
  \includegraphics[width=1\textwidth]{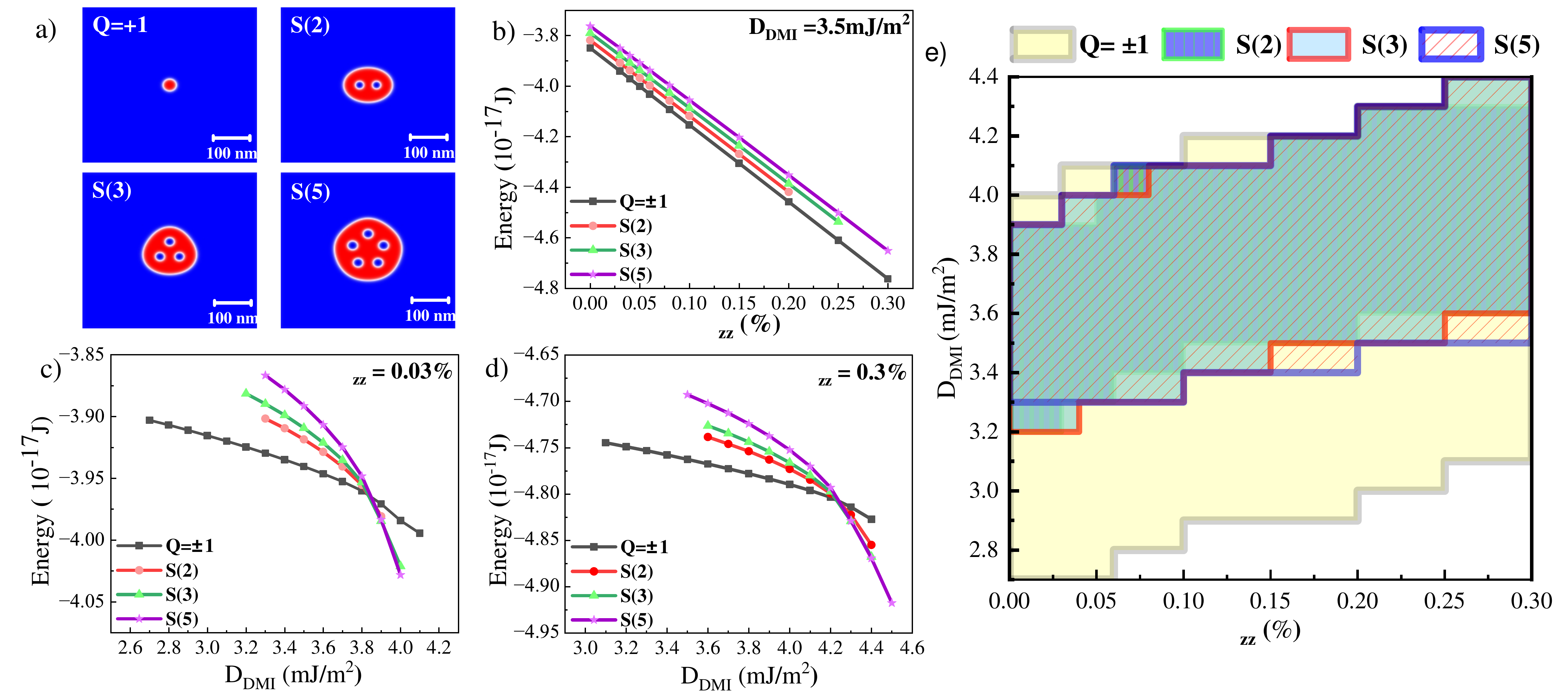}
  \caption {a) Magnetization distribution for stable skyrmion and skyrmion bags in magnetic thin film at DMI = $3.5 \ mJ/m^2$ b) Total energy vs out-of-plane strain ($ \epsilon_{zz}$ ) applied at a fixed DMI = $3.5 \ mJ/m^2$ for different topological charges. c) Total energy vs DMI for $ \epsilon_{zz}$  = $0.03\%$ d) Total energy vs DMI for $ \epsilon_{zz}$ = $0.3\%$. e) Phase diagram showing the stability window of skyrmion bags as a function of wide range of DMI and  $ \epsilon_{zz}$ values.} 
  \label{fig1}
\end{figure*}
We used following parameters \cite{43,66} for the simulation of all the topological charges (TC's): exchange constant $A_{ex}$ = 15 pJ/m, perpendicular anisotropy constant K = 0.8$MJ/m^3$ , Dzyaloshinskii–Moriya interaction $D_{DMI}$ = 3.5 $mJ/m^2$  , saturation magnetization $M_{sat} = 580 \ KA/m$, spin polarization P = 0.4 and current density $\textbf{j}$ = $-5\times 10^{10} \ A/m^2$. To highlight the effect of strain gradient on the dynamics, we purposefully chose $\alpha$ and $\beta$ to have the same value i.e., 0.03. Under this condition, in the absence of strain gradient the skyrmion or skyrmion bags, moves in a straight line without showing any Skyrmion Hall angle. Thus any deflection in the current induced motion will be an effect of forces originating due the strain gradient. 
\section{Result and Discussion}
\subsection{Strain induced stability of Topological Charges}
We began by investigating the stabilization of different topological charges, such as skyrmions and skyrmion bags, in the film under zero strain, using the initial magnetization and energy parameters outlined in the previous section. FIG.  \ref{fig1}(a) displays the out-of-plane magnetization component for skyrmions (Q = +1) and skyrmion bags labeled S(2), S(3), and S(5), where the outer skyrmion with Q = +1 contains two, three, and five inner skyrmions, respectively, each with Q = -1. Due to the skyrmion-skyrmion repulsion \cite{1,69}, the bag radius increases with the number of inner skyrmions hosted within the outer skyrmion boundary.

FIG.  \ref{fig1}(b) shows how the total magnetic energy evolves for the different magnetization configurations  as a function of out-of-plane strain $(\epsilon_{zz})$ in the film, with the Dzyaloshinskii-Moriya interaction (DMI) set at $D_{DMI} =$ 3.5 $mJ/m^2$.
For clarity, the data for each spin configuration has been vertically shifted upwards by an equal amount of $E=3\times 10^{-19} J$ relative to one another. The total energy decreases with increasing $ \epsilon_{zz}$ showing that the magnetic configurations are more stable for higher values of $ \epsilon_{zz}$ in the film. It should also be noted that the range for $ \epsilon_{zz}$  in which the skyrmion bags are stable configuration depends upon the total number of inner skyrmions (N) i.e., higher the N, larger is the range for $ \epsilon_{zz}$ in which S(N) is stable. For example, at $D_{DMI} =$ 3.5  $mJ/m^2$, S (2) is stable as long as $ \epsilon_{zz}  \leq 0.20 \%$ while S (5) is stable for  $\epsilon_{zz}  \leq 0.30 \%$ . We consider a spin configuration to be stable as long as the shape is not deformed and its symmetry is preserved with respect to the initialized configuration shown in Fig.1(a). The stability range for $ \epsilon_{zz}$ also depends upon the value of DMI. FIG.  \ref{fig1}(c) and FIG.  \ref{fig1}(d) shows the evolution of the total energy as a function of DMI for two different out of plane strain values of $ \epsilon_{zz} = 0.03 \%$   and $ \epsilon_{zz} = 0.3 \%$, respectively. It can be seen that for $\epsilon_{zz}  = 0.03 \%$, skyrmion is the most stable configuration for $D_{DMI} \leq 3.8$. We define critical DMI $(D^c_{DMI})$ as the DMI below which skyrmion (Q = ±1) is more stable than the S(N). We also note that below $(D^c_{DMI})$, the stability of the skyrmion bags decreases as the number of inner skyrmions increases. Conversely, when the DMI exceeds $(D^c_{DMI})$, larger skyrmion bags with more inner skyrmions exhibit greater stability compared to smaller bags. Furthermore, it was observed that $D^c_{DMI}$ shifts towards higher value of 4.2 $mJ/m^2$ (FIG.  \ref{fig1}(d)) as $ \epsilon_{zz}$ is increased to $ 0.3 \%$. These trend can also be observed in the magnetic phase diagram for skyrmion and skyrmion bags as function of $D_{DMI}$ and $ \epsilon_{zz}$ shown in FIG.  \ref{fig1}(e).

\subsection{\texorpdfstring{Dynamics of Skyrmion $Q = -1$}{Dynamics of Skyrmion Q = -1}}

\begin{figure}[h]
\centering 
\includegraphics[width=0.5\textwidth,height=8.3cm]{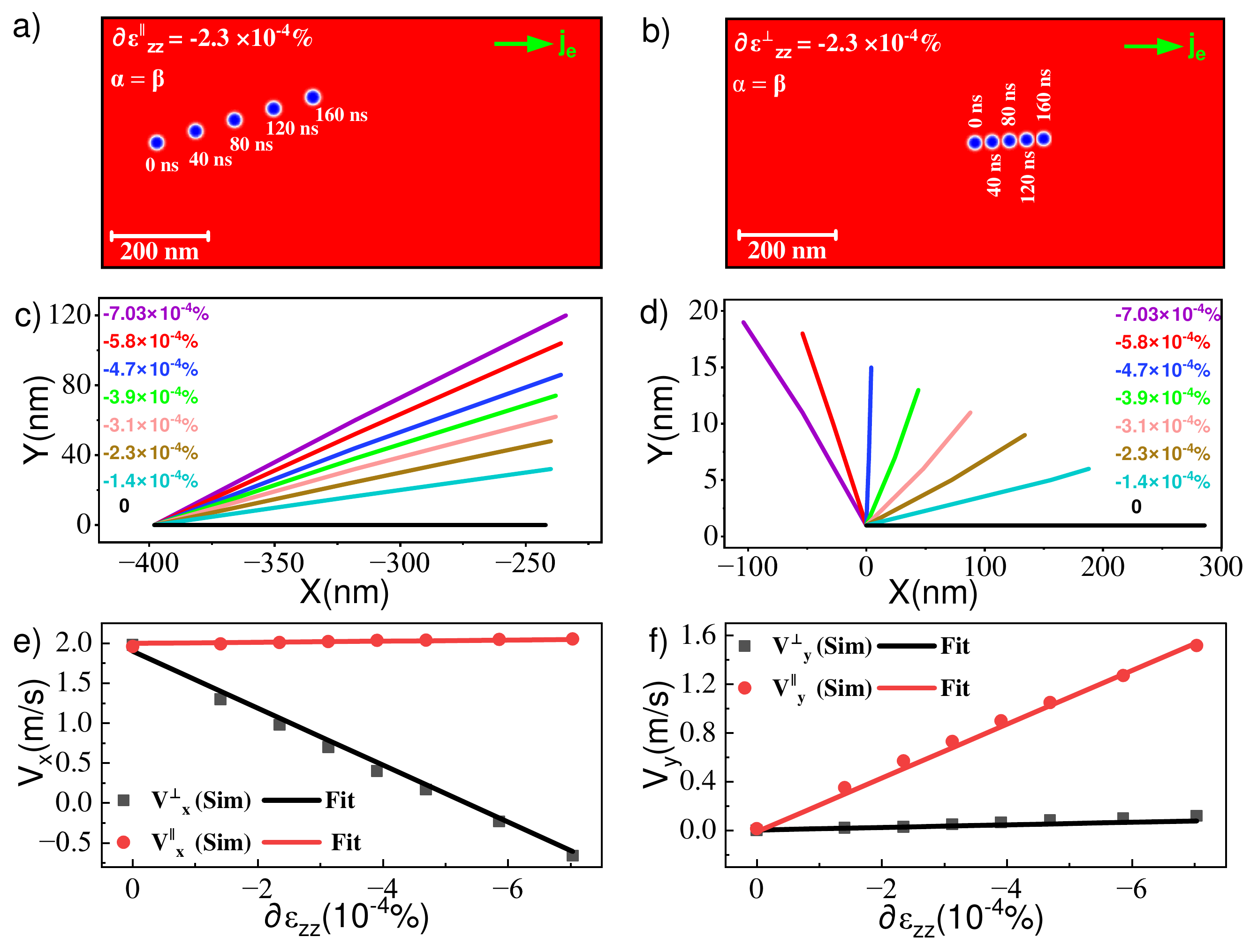}	
\caption{ Current induced Dynamics of Q = -1 skyrmion for a fixed out-of-plane strain gradient ( $\partial \epsilon_{zz}$) applied (a) parallel and (b) perpendicular to the electron flow. The green arrow marks the direction of electron flow. Trajectories of skyrmion for different  $\partial \epsilon_{zz}$ applied (c) parallel and (d) perpendicular to the electron flow. (e) The x-component of velocity ($V_x$)  and (f)   y-component of velocity ($V_y$) for different strain gradient applied parallel (solid red circle symbols)  and perpendicular (solid black square symbols) to electron flow. The solid lines in (e) and (f) are fit to the simulated data using Thiele equation.} 
	\label{fig2}
\end{figure}

Now we evaluate the current induced dynamics of skyrmion in the presence of strain gradient. Firstly the out-of-plane strain gradient is chosen to be along the direction of electron flow (\textbf{j}$_\textbf{e}$ shown by green arrow at the
upper-right corner of \ref{fig2}(a,b)). Here $Q=-1$ skyrmion is stabilized in the left-center of the nanotrack of dimensions $1024 \times 512 \times 0.4 nm^3$ in the presence of negative strain gradient $(\partial \epsilon_{zz}=-2.3\times 10^{-4}\%)$ along x-direction. The dynamics of skyrmion for the current-in-plane (CIP) geometry is governed by the Landau-Lifshitz-Gilbert (LLG) equation shown in equation (3). To ensure that the impact of the boundary potential on the dynamics of the skyrmions is negligible, we consider dynamics far from the edges of the nanotrack. This allowed us to focus solely on the effect of strain gradient on the dynamics of skyrmion.

FIG.  \ref{fig2}(a) shows the current induced dynamics of $Q=-1$ skyrmion when the strain gradient was fixed at  $\partial \epsilon_{zz}^\parallel=-2.3\times10^{-4}\%$. The skyrmion is seen to be deflected towards the +y direction. It should be noted that as $\alpha$ = $\beta$, the deflection observed above originates completely from the force acting on the skyrmion due to the presence of strain gradient. FIG.  \ref{fig2}(c) shows the trajectories of skyrmions $Q=-1$ under different values of negative strain gradient applied along the direction of \textbf{j}$_\textbf{e}$. 
It is clear that, at zero strain skyrmion is seen to move in a straight line (black curve in FIG.  \ref{fig2}(c)) and with increasing strain gradient the deflection along the y-direction is enhanced. In FIG.  \ref{fig2}(e,f), the solid red circle symbols represents the $V_x^\parallel$ and $V_y^\parallel$  defined as velocity component along x and y direction respectively, when the strain gradient is  applied parallel to the direction of \textbf{j}$_\textbf{e}$. Our simulations showed that  $V_x^\parallel$  practically stays constant, however $V_y^\parallel$  continues to rise as we increase the negative strain gradient (solid red circle symbol and solid red line). \\
\ \ FIG.  \ref{fig2}(b,d) shows the simulation results when the negative out of plane strain gradient $(\partial \epsilon_{zz}^\perp=-2.3\times 10^{-4}\%)$ was applied along y-direction, i.e., perpendicular to the direction of \textbf{j}$_\textbf{e}$. In this case the deflection is along the +y direction and the deflection increases with the increasing strain gradient. \\
In FIG. \ref{fig2}(e,f), we analyze the behavior of the velocity components $V_x^\perp$ and $V_y^\perp$ in response to a strain gradient applied perpendicular to  \textbf{j}$_\textbf{e}$. Notably, $V_y^\perp$ remains nearly constant across varying strain gradient values, indicating that the y-component of the velocity is less sensitive to these changes. Conversely, the x-component $V_x^\perp$ displays a significant trend: it initially decreases with increasing strain, demonstrating a negative slope. This decrease continues until the strain gradient reaches approximately 
$\partial \epsilon_{zz}^\perp{=}-4.7\times10^{-4}\%$, at which point the behavior shifts, and $V_x^\perp$ changes sign. This indicates that the skyrmion begins to move in the direction opposite to the electron flow \textbf{j}$_\textbf{e}$ , as illustrated by the solid black square symbols and the black line in the figure. These results highlight that strain gradient perpendicular or parallel to the electron flow can be used to control the dynamics of skyrmion very effectively and could be a key ingredient for the skyrmion-based memory application.\\
The observed trajectory of skyrmion can be explained by the following Thiele Equation \cite{70},
\begin{equation}
\textbf G \times (\textbf v_e -\textbf v) + \textbf D(\beta \textbf v_e-\alpha \textbf v )+ \textbf F_{strain}=0 
\end{equation}
where $ \textbf G=\frac{4\pi QM_{sat} t}{\gamma}$ is the gyromagnetic coupling vector, Q is topological charge, $\gamma$ is gyromagnetic ratio, $ t $ is thickness of thin film, $\textbf D$ is the dissipative tensor, $\alpha$ is the gilbert damping constant, $\beta$ is the non-adiabatic spin transfer parameter, $v_e=\mu_B g P \textbf{j}_\textbf{e}/ 2e M_{sat} $ is the velocity of conduction electron, P is spin current polarization, $M_{sat} $ is saturation magnetization, \textbf j is current density , $\textbf v$ is the velocity of topological charge. The first term describes the Magnus force, $2^{nd}$ term is the dissipative force and the third term  $\textbf F = -\nabla W$  originates due the applied strain gradient. It should be noted that as we were interested in dynamics far from the boundary of the nanotrack, the force due to the boundary potential is neglected here.

The dissipative tensor components are calculated as:
\begin{equation}
  D_{ij}= \frac{M_{sat} L}{ 4 \pi \gamma} \int \partial _i \textbf m \cdot \partial_j \textbf m \ d^2r \nonumber  
\end{equation}
For a circular Néel Skyrmion, $D_{xx} = D_{yy} = D$ and $D_{xy} = D_{yx} = 0$, 
which was also verified by carrying out the above integration over the region containing the spin texture. The dissipative tensor \textbf{D} now has the following form,
$$\textbf D=  \left( \begin{array}{cc}
  D & 0 \\
    0 & D \\  
\end{array}\right) $$Thus, the motion described by the Thiele equation (4), leads to the following expression for the velocity components,
	\begin{equation}	
    v_x= \frac{(G^2+D^2 \alpha \beta) v_e}{G^2+\alpha^2  D^2 } + \frac{GF_y+\alpha DF_x }{G^2+\alpha^2  D^2} 			\end{equation}        
	\begin{equation}		 v_y{=} \frac {-GF_x  + \alpha DF_y}{G^2+\alpha^2  D^2 }+\frac{DG(\alpha-\beta) v_e}{G^2+\alpha^2  D^2}		\end{equation}For $\alpha {=}\beta$, the second term in equation (6) will vanish and the equations will then depict the effect of strain gradient solely. Moreover, we can consider $F_y$ and $F_x$ to be negligible when the strain is applied parallel and perpendicular to \textbf{j}$_\textbf{e}$ respectively. Finally, $\alpha$ being very small and  assuming $D \equiv G$ for skyrmion, the above two equations for the velocity components gets modified as
\begin{equation}	v_x^\parallel{=}v_e +\frac{\alpha F_x }{|G|}  \ \ and  \ \ v_y^\parallel{=} -\frac {F_x }{G }	 	\end{equation}	       
\begin{equation}	           v_x^\perp{=}v_e+\frac{F_y } {G}     \ \ and \ \ v_y^\perp {=} \frac{\alpha F_y }{|G| }	             \end{equation}   The solid line in FIG.  \ref{fig2}(e,f) represents the fit to the simulated points(solid symbols) based on equation (7,8).  
 $v_x^\parallel$ for skyrmion Q = -1  remains nearly constant (solid red circle and solid red line in FIG.  \ref{fig2}(e)) with the application of strain gradient since the second term in Equation (7) contributes insignificantly due to the small value of $\alpha$, allowing it to move in the direction of \textbf{j}$_\textbf{e}$. However, $v_y^\parallel$ increases linearly as the force due to the strain gradient  pushes the Q = -1 skyrmion upward shown in FIG.  \ref{fig2}(f). 
 
When the strain gradient is applied perpendicular to electron flow, $v_x^\perp$ increases linearly with a negative slope ( solid black square and solid black line in FIG.  \ref{fig2}(e)). With increasing strain gradient, the $2^{nd}$ term in $v_x^\perp$ (Equation (8)) starts to dominate and finally will lead to movement of Q = -1 skyrmion in a direction opposite to the motion of  electron flow. 
Similar current induced dynamics were obtained for Q = +1 skyrmions (data not shown) under the out of plane strain gradient.

\subsection{Killing of Skyrmion Hall Effect (SkHE)}

\begin{figure}[h]
     \includegraphics[width=0.485\textwidth]{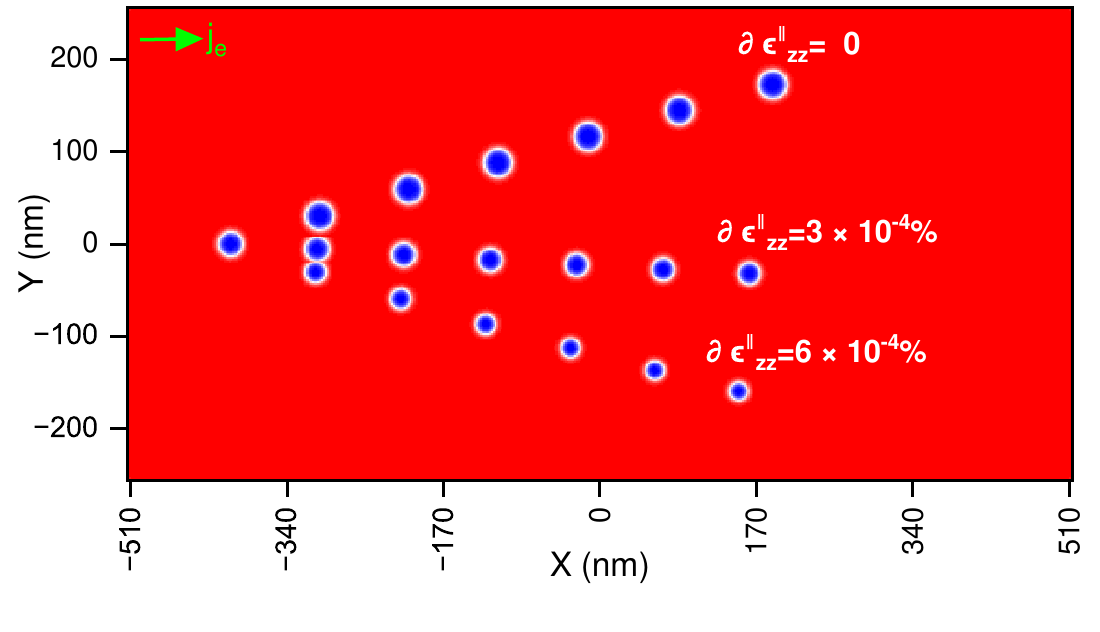}
	\caption{Killing of skyrmion hall effect with the application of out of plane strain gradient. Trajectory of  skyrmion Q = -1 when $\alpha \neq \beta $ at different positive strain gradient applied parallel to direction of  electron flow. Green arrow at the upper-left corner marks the direction of electron flow.} 
	\label{fig3}
\end{figure}
The Skyrmion Hall effect, a phenomenon where skyrmions drift at an angle when subjected to a current, poses a significant challenge for memory applications. Here we demonstrate that applying a strain gradient parallel to the direction of \textbf{j}$_\textbf{e}$ can effectively eliminate the Skyrmion Hall effect. We first simulate the dynamics for the Q = -1 skyrmion in the absence of strain gradient ($\partial \epsilon_{zz}=0$) and consider $\alpha \neq \beta$. The snapshots  of the skyrmion trajectory is shown in FIG.  \ref{fig3}. In the absence of strain gradient, the skyrmion is deflected in the positive y-direction in accordance with the SkHE. With the application of positive strain gradient $\partial \epsilon_{zz}= 6 \times 10^{-4}\%$, the skyrmion is forced to deflect in the direction opposite to that of conventional SkHE, and thus the strength of the strain gradient can be effectively used to compensate SkHE. In FIG.  \ref{fig3} it is shown that for $\partial \epsilon_{zz}= 3 \times 10^{-4}\%$, the skyrmion will move in a straight path, thus demonstrating killing of SkHE.

\subsection{Dynamics of  Skyrmion Bags}

\begin{figure}[h]
\centering
\includegraphics[width=0.5\textwidth,]{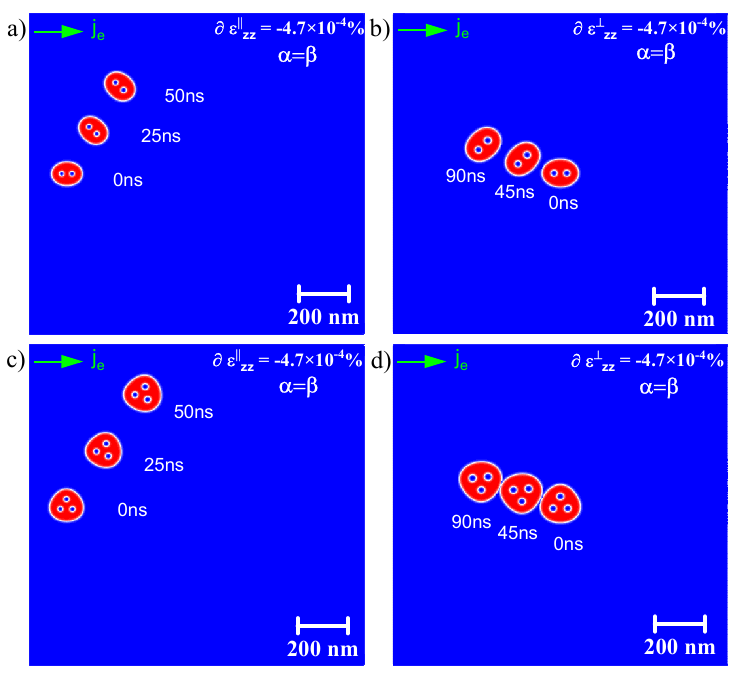}
\caption{ Current induced dynamics of skyrmion bags in the presence of strain gradient. (a,c) Trajectory of S (2) and S (3) when negative strain gradient $\partial \epsilon_{zz}= -4.7 \times 10^{-4}\%$ is applied in the film in the direction parallel to electron flow. (b,d) Trajectory of S (2) and S (3) when negative strain gradient $\partial \epsilon_{zz}= -4.7 \times 10^{-4}\%$ is applied in the film in the direction perpendicular to current induced. Green arrow at the upper-left corner shows the direction of  electron flow.}
\label{fig4}
\end{figure}

For simulating current induced dynamics of skyrmion bags in the presence of out of plane strain gradient we consider a magnetic film of dimension $1024 \times 1024 \times 0.4 nm^3$ and monitored current induced dynamics far from edges of the film. When strain gradient is applied parallel to direction of electron flow (\textbf{j}$_\textbf{e}$), the skyrmion bag is initialized on left-center of the film and for strain gradient perpendicular to \textbf{j}$_\textbf{e}$, bags are stabilized at the center of the film. Here again we consider $\alpha = \beta$ to study the effect of strain gradient solely. FIG.  \ref{fig4} shows the snapshots for S(2) and S(3) when the strain gradient $(\partial \epsilon_{zz} {=}-4.7\times 10^{-4}\%)$ is applied parallel ( FIG.  \ref{fig4}(a,c))  and perpendicular ( FIG.  \ref{fig4}(b,d)) to the \textbf{j}$_\textbf{e}$. The forces due to the strain gradient causes deflection of topological charges from  straight path. Moreover for strain gradient perpendicular to the electron flow, we observe motion of the bags opposite to the direction of \textbf{j}$_\textbf{e}$.

\begin{figure}[h]
\centering
\includegraphics[width=0.48\textwidth,]{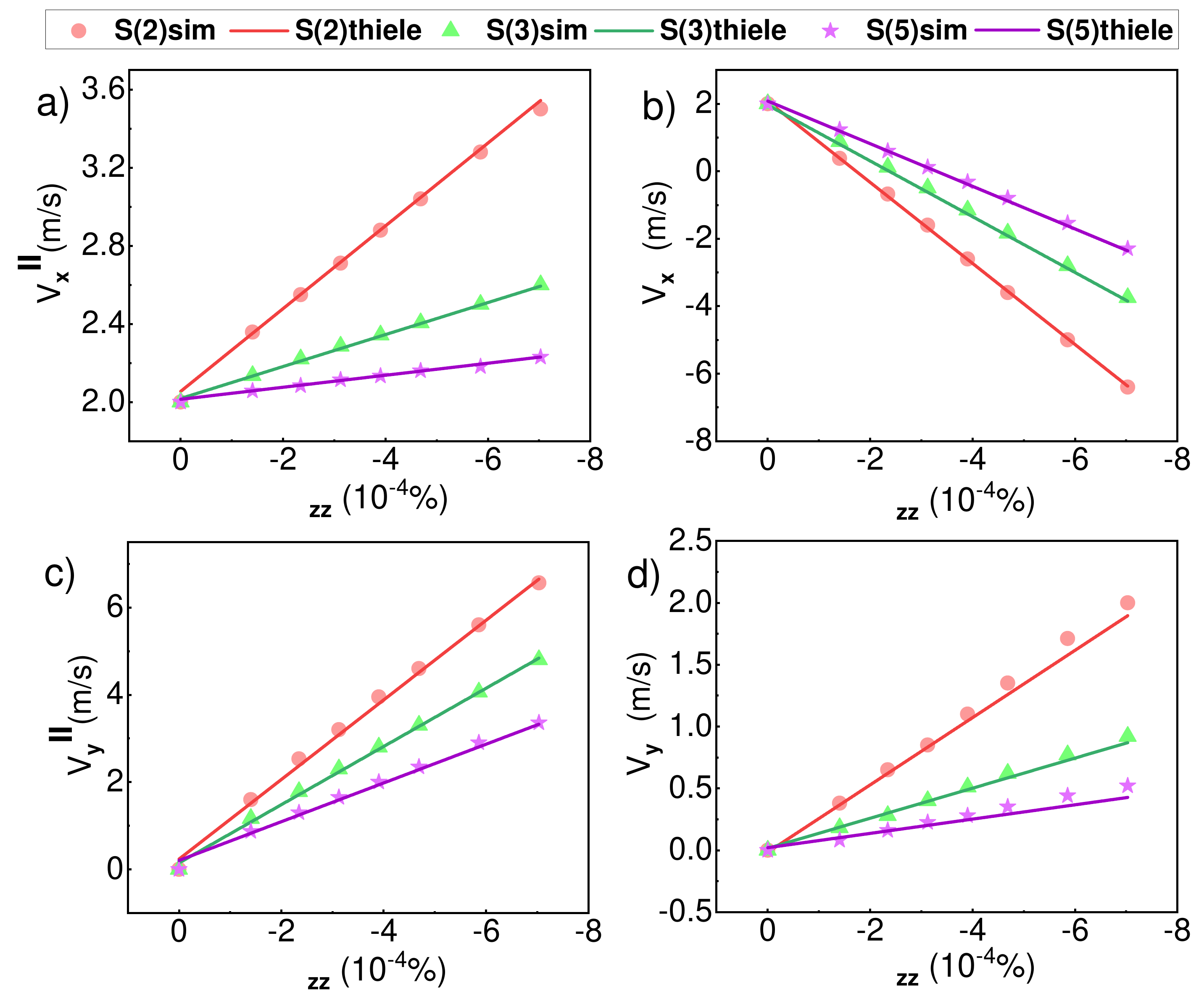}
\caption{(a,c) Velocity component Vx and Vy vs negative strain gradient when applied parallel for different skyrmion bags. (b,d) Velocity component Vx and Vy vs negative strain gradient when applied perpendicular for different skyrmion bags. Solid lines are fit to the simulated results (symbols) using Thiele equation.}
\label{fig5}
\end{figure}
The dynamics of skyrmion bags were investigated under varying strain gradients.  FIG.  \ref{fig5} illustrates the velocity components derived from the motion of skyrmion bags at different strain gradient values. It is evident that the velocities are influenced by the topological charges: as the topological charge increases, the velocity decreases for a given strain gradient.
$v_x^\parallel$  and $v_y^\parallel$ , representing x and y component of velocities when strain gradient ($\partial \epsilon_{zz}$) is applied parallel to the \textbf{j}$_\textbf{e}$ is shown in FIG.  \ref{fig5}(a,c). We observe that  $v_x^\parallel$  for any particular topological charge [S (2), S (3), S (5)] shows a very slow but linear increase with the $\partial \epsilon_{zz}$ which is in contrast of  $v_y^\parallel$ where velocity increases sharply with increasing $\partial \epsilon_{zz}$. In FIG.  \ref{fig5}(b,d), we plot the $v_x^\perp$  and $v_y^\perp$, representing the  x and y component of velocity when strain gradient ($\partial \epsilon_{zz}$) is applied perpendicular to the \textbf{j}$_\textbf{e}$. In this scenario, the force induced by the strain gradient causes the skyrmions bags to move in a direction opposite to  \textbf{j}$_\textbf{e}$. This is reflected in the behavior of $v_x^\perp$  as a function of the strain gradient, which shows a negative slope. This negative slope emphasizes that at higher strain gradient values, the motion of the skyrmions is indeed reversed, moving against the direction of  \textbf{j}$_\textbf{e}$. The $v_y^\perp$ for skyrmion bags  increases linearly with increasing $\partial \epsilon_{zz}$,  as evident from a positive slope shown in FIG.\ref{fig5}(d).
To explain the dynamics of skyrmion bag we once again consider the Thiele Equation in the absence of boundary potential which is given as:
\begin{equation}
\textbf G \times (\textbf v_e -\textbf v) + \textbf D(\beta \textbf v_e-\alpha \textbf v )+ \textbf F_{strain}{=}0 \nonumber
\end{equation}
 Since skyrmion bags in general does not have to be axially symmetric, hence the dissipative tensor $\textbf D$ can have all non-zero components, $$\textbf D=  \left( \begin{array}{cc}
  D_{xx} & D_{xy} \\
    D_{yx} & D_{yy} \\  
\end{array}\right) $$
For each skyrmion bag we evaluate all the components of the Dissipative tensor ( $D_{xx}, D_{yy},D_{xy} \ and \ D_{yx}$) using the integration defined earlier in text.
By Solving the equation (4) and incorporating above expression of $\textbf D$ and considering $\alpha = \beta$, we get the expression for parallel and perpendicular velocity components as,

\begin{equation}  v_x^\parallel{=}v_e +\frac{\alpha F_x D_{yy}}{G^2+\alpha^2 D_{xx} D_{yy}-\alpha^2 D_{xy} D_{yx}}   \end{equation}
\begin{equation}      v_y^\parallel{=} -\frac {F_x(G^2-\alpha^2 D_{xy} D_{yx}) }{(G-\alpha D_{xy})(G^2+\alpha^2 D_{xx} D_{yy}-\alpha^2 D_{xy} D_{yx}) }  \end{equation}      
\begin{equation}           v_x^\perp{=}v_e+\frac{F_y(G-\alpha D_{xy}) } {G^2+\alpha^2 D_{xx} D_{yy}-\alpha^2 D_{xy} D_{yx}}      \end{equation}

\begin{equation}  v_y^\perp = \frac{\alpha F_y D_{xx}}{G^2+\alpha^2 D_{xx} D_{yy}-\alpha^2 D_{xy} D_{yx} }              \end{equation}    
These equation clearly shows that the velocity of the bags are determined not only by the topological charge number Q but also by the force originating due to the strain gradient.
In FIG.  \ref{fig5}, the solid lines are the fit to the simulated results using the above mentioned equations (9–12). The Thiele equation closely matches the simulated results for skyrmion bags.
For $\partial \epsilon_{zz}$ applied perpendicular to the \textbf{j}$_\textbf{e}$, with increasing strain gradient, $2^{nd}$ term in equation (11) starts to dominate and thus the bags will move in direction opposite to the electron flow.

\section{\textbf{Conclusion}}

In summary, out-of-plane strain and its gradient significantly influence both the stability and dynamics of topological charges in ferromagnetic films, presenting opportunities for advancements in next-generation memory and spintronic devices.  Out-of-plane strain enhances the stability of topological charges. In the presence of out of plane strain, below a certain critical Dzyaloshinskii-Moriya interaction $(D^c_{DMI})$ value, skyrmions are found to be more stable. However, when the $D_{DMI}$ exceeds this critical threshold, configurations with higher topological charges (S(N) with higher N) become more stable.  We further studied the current induced dynamics of topological charges in the presence of strain gradient applied parallel or perpendicular to the electron flow (\textbf{j}$_\textbf{e}$). The strain gradient not only effects the skyrmion velocity but also determine the deflection of skyrmion as it moves along the gradient. The deflection is larger for higher strain gradients and can be used to kill skyrmion hall effect. We also demonstrated the effect of strain gradient on the dynamics of skyrmion bags. The velocity of skyrmion bags is influenced by their topological charge: higher topological charges correspond to lower velocities for the same spin current density and strain gradient. With increasing strain gradient, the x and y component of the velocity for the skyrmion bags increases linearly demonstrating strain gradient could be an efficient tool to control dynamics of skyrmion bags in the racetrack memory. The simulation results align well with the calculations derived from the Thiele equation, indicating a strong consistency between the two methods. This agreement reinforces the validity of the theoretical framework used to analyse the dynamics of topological charges.

\section*{Acknowledgement} \vspace{-\baselineskip}
This work was financially supported by Science and Engineering Research Board (SERB) Start-Up Research Grant (SRG/2021/002186) and Scheme for Translational and Advanced Research in Science (STARS) ( MoE-STARS/STARS-2/2023-0437).

\section*{Author Declaration} 
\textbf{Conflict of Interest}: The authors have no conflicts to disclose.\vspace{-\baselineskip}

\section*{Data Availability} \vspace{-\baselineskip}
The data that support the findings of this study are available from the corresponding author upon reasonable request.

\bibliographystyle{apsrev4-2}
\bibliography{ref}

 \end{document}